\documentclass[twocolumn,showpacs,prb,amsmath,amssymb]{revtex4}
\usepackage{color}
\usepackage{times}
\usepackage{float}
\usepackage{flafter}
\usepackage{epsfig}%

\newcommand{\gammamct}{\gamma_{\text{MCT}}}
\newcommand{\gammadot}{\dot{\gamma}}

\newcommand{\etaapp}{\eta_{\text {app}}}

\newcommand{\rcab}{r_{\text{c},\alpha\beta}}
\newcommand{\sigmay}{\sigma_{\text{yield}}}

\newcommand{\Tc}{T_{\text c}}
\newcommand{\Fe}{F^{\text e}}

\newcommand{\tauLJ}{\tau_{\text LJ}}

\newcommand{\rhoA}{\rho_{\text A}}
\newcommand{\rhoB}{\rho_{\text B}}

\newcommand{\NB}{N_{\text B}}
\newcommand{\NA}{N_{\text A}}
\newcommand{\myeq}{\!=\!}

\newcommand{\tw}{t_{\text {w}}}
\newcommand{\kB}{k_{\text {B}}}

\newcommand{\add}[1]{\textcolor{black}{#1}}
\newcommand{\remove}[1]{}

\begin{document}

\large

%\preprint{Submitted to J. Chem. Physics}
%\twocolumn[\hsize\textwidth\columnwidth\hsize\csname@twocolumnfalse\endcsname

\title{Structural relaxation and rheological response
  of a driven amorphous system}
\author{F. Varnik}

%\author{F. Varnik$^{(1,2)}$, O. Bocquet$^{(2)}$, J.-L. Barrat$^{(2)}$}
\address{Max-Planck Institut f\"ur Eisenforschung, Max-Planck Stra{\ss}e 1,
40237 D\"usseldorf,
Germany}
%(2) Laboratoire de Physique de la Mati\`ere Condens\'e et Nanostructures,
%Universit\'e Lyon I and CNRS, 69622 Villeurbanne Cedex, France\\
%}
\date{\today}% It is always \today, today, but any date may be explicitly specified

%xxxx Bei Franosch alle Autoren nennen.
%xxxxx Achsebeschriftung Fig 7a

\begin{abstract}
The interplay between the structural relaxation and the rheological
response of a simple amorphous system 
(an 80:20 binary Lennard-Jones mixture 
[W. Kob and H.C. Andersen, PRL {\bf 73}, 1376 (1994)]
 is studied via molecular dynamics simulations.
In the quiescent state, the model
is well known for its sluggish dynamics and a two step relaxation of correlation
functions at low temperatures. An ideal glass transition temperature of
$\Tc \myeq 0.435$ has been identified in the previous studies via the analysis of
the system's dynamics in the frame work of the mode coupling theory of the
glass transition [W. Kob and H.C. Andersen, PRE {\bf 51}, 4626 (1995)]. 
In the present work, we focus on the question whether a signature of this
ideal glass transition can also be found in the case where the system's
dynamics is driven by a shear motion. Indeed, the following distinction
in the structural relaxation is found: In the supercooled state, the
structural relaxation is dominated by the shear at relatively high shear
rates, $\gammadot$, whereas at sufficiently low $\gammadot$ the (shear-independent)
equilibrium relaxation is recovered. In contrast to this, the
structural relaxation of a \emph{glass} is always driven  by shear.
This distinct
behavior of the correlation functions is also reflected in the
rheological response. In the supercooled state, the shear viscosity,
$\eta$, decreases with increasing shear rate (shear thinning) at high shear rates,
 but then converges toward a constant as the $\gammadot$ is decreased below
a (temperature-dependent) threshold value. Below $\Tc$, on the other
hand, the shear viscosity grows as $\eta \propto 1/\gammadot$
suggesting a divergence at $\gammadot \myeq 0$. Thus, 
within the accessible observation time window, a transition toward a 
non-ergodic state seems to occur in the driven glass as the driving 
force approaches zero. As to the flow curves (stress versus shear rate), a plateau forms at low
shear rates in the glassy phase. \add{A consequence of this stress plateau for
Poiseuille-type flows is demonstrated.}
\end{abstract}
% PACS, the Physics and Astronomy Classification Scheme:
%%05.70.Ln   Non-equilibrium and irreversible thermodynamics (see also 82.40.Bj Oscillations, chaos, and bifurcations in physical chemistry and chemical physics)
%%64.70.Pf   Glass transitions
%%83.60.Fg Shear rate dependent viscosity
\maketitle

%%%%%%%%%%%%%%%%%%%%%%%%%%%%%%%%%%%%%%%%%%%
\section{introduction}
\label{section:introduction}%%%%%%%%%%%%%%%
Although simple at first sight, suspensions of spherical colloidal
particles under shear exhibit a rich phenomenology. In the dilute
regime, at temperatures corresponding to the liquid state, forced
Rayleigh scattering experiments \cite{Qiu-PRL61-1988} show an increase of
diffusion constant upon shearing (shear thinning), distinct from Taylor
dispersion \cite{Taylor-ProcRSocLondonA-219::1953}. Indeed, Taylor dispersion
(displacement of particles along the flow direction, $x$, as they move in the
direction of shear gradient) would give rise to an enhancement of mobility
via $\langle\Delta x^2(t)\rangle\sim t^3$, whereas shear thinning leads to
$ \langle\Delta x^2(t)\rangle  \myeq  D(\gammadot) t$,  with a  diffusion coefficient,
$D(\gammadot)$, which increases with increasing $\gammadot$ in a non-trivial fashion.

When shearing starts in the crystalline phase, however, shear
induced melting of the crystalline structure is observed both in
light scattering experiments
\cite{Ackerson-Clark::PRL46-1981,Ackerson::JRheol34-1990} as well as
in Brownian dynamics simulations \cite{Xue-Grest::PRL64::1990}.
Furthermore, in the shear melted regime, experiments show evidence
for shear thinning due to the presence of freely slipping two
dimensional crystalline layers
\cite{Ackerson-Clark::PRL46-1981,Ackerson::JRheol34-1990}.
Simulations also show a shear thinning regime below a "critical"
shear rate, $\dot{\gamma}_{\text c}$, followed by a transition to a
string-like order for $\gammadot>\dot{\gamma}_{\text c}$
\cite{Xue-Grest::PRL64::1990}.

On the other hand, studies of disordered suspensions of hard spheres
show that shear thinning and shear melting phenomena may also occur
in the absence of a crystalline structure
\cite{Laun-et-al::JRheol36-1992,Petekidis::Physica-A306-2002,Petekidis::PRE66-2002,
Petekidis-Vlassopoulos-Pusey::FaradayDiscuss::2003}. Similar
observations have also been made in light scattering echo studies of
(disordered) dense emulsions
\cite{Hebraud-Lequeux-Munch::PRL78-1997}. Brownian dynamics
simulations show that shear thinning in concentrated colloidal
suspensions is related to the fact that, in the limit of low shear
rates, the main contribution to the shear stress originates from the
Brownian motion of colloidal particles and that this contribution
decreases with $\gammadot$
\cite{Phung-Brady-Bossis-JFluidMech::1996}.

Many theoretical approaches have been proposed in recent years
aiming at a description of the rheology of disordered materials.
Some examples are studies of driven $p$-spin glasses
\cite{Berthier-Barrat-Kurchan::PRE}, the so called ``soft glassy
rheology'' model (SGR)  \cite{Sollich::PRE-1998} and non-equilibrium
extensions of the mode coupling theory (MCT) of the glass transition
\cite{Fuchs-Cates::PRL-2002,Fuchs-Cates::FaradayDiscuss-2003,Miyazaki-Reichman-PRE66,Miyazaki-Reichman-Yamamoto-PRE70}.

These approaches have the common feature of predicting (or, more precisely,
``reproducing'', as the phenomenon was known before the theories have been developed)
the shear thinning phenomenon in an \emph{ergodic} disordered system, i.e.\ in
a system where temporal correlations decay to zero in a finite time. In these
systems, shear thinning is commonly attributed to the competition between
a typical structural relaxation time (of the quiescent system) and the time
scale imposed by the external drive. A shear thinning behavior is expected as soon as
the latter becomes significantly shorter than the former, in which case the
structural relaxation is said to be driven by the external force \cite{Berthier-Barrat-Kurchan::PRE}.
This idea is also supported by the observation that even a simple liquid may exhibit shear thinning
provided that the shear rate is high enough in order for shear disturbance to dominate the
system dynamics~\cite{Evans-Morriss::StatMechNonEqLiq1990}.

However, the statements of different theories diversify as it comes
to the rheological response of a ``jammed system'' such as an
amorphous solid. The discrepancy is best seen when considering the
behavior of the yield stress, i.e.\ the stress response in the limit
of vanishing shear rate (driving force), $\sigmay \equiv
\sigma(\gammadot \to 0)$. While calculations based on the $p$-spin
glasses \add{in the thermodynamic limit} predict that no yield stress 
exists in the system \cite{Berthier-Barrat-Kurchan::PRE} 
($\sigmay$ vanishes with a power law, $\sigmay\propto \gammadot^{\alpha}$ with
$0<\alpha<1$), the SGR model \cite{Sollich::PRE-1998} predicts the
onset of a (dynamic) yield stress at the ``jamming transition'',
$\sigmay \myeq 1-x$. Here, $x$ is a  noise temperature which
controls the "degree of jamming" or the distance from the glass
transition.  $x \myeq 1$ corresponds to the glass transition (or
``jamming'') temperature, and $0<x<1$ characterizes the glassy or
``jammed'' phase (for noise temperatures above the transition,
$1<x<2$, a power law decrease of the shear viscosity with the
applied shear rate is found \cite{Sollich::PRE-1998}).

Interestingly, the presence of a finite yield stress in an amorphous
solid has also been predicted within the simplest (or idealized)
version of the non-equilibrium MCT approach
\cite{Fuchs-Cates::FaradayDiscuss-2003,Fuchs-Cates::PRL-2002}.
A related MCT approach to the fluctuations around the steady state
has recently been proposed in reference \cite{Miyazaki-Reichman-PRE66,Miyazaki-Reichman-Yamamoto-PRE70}.
The issue of yield stress, however, could not be
addressed in that approach.

Despite the above mentioned qualitatively different predictions on
the stress response of a glass, both the $p$-spin approach and the
non-equilibrium MCT commonly predict that the presence of an
external drive leads to a \emph{melting} of the amorphous solid.

\add{In reference \cite{Berthier-Barrat-Kurchan::PRE} it is argued
that the main reason for the absence of a yield stress in the
$p$-spin model lies in its mean-field nature so that free energy
barriers are impenetrable. If the system is quenched from a high
temperature to a temperature in the glassy phase while at the same
time being driven by a nonconservative force, it remains sliding
above the free energy \emph{threshold} below which the free energy
surface is split into exponentially many disconnected regions
\cite{Berthier-Barrat-Kurchan::PRE}. This threshold does not exist
at high temperatures but emerges at a critical temperature, $\Tc$,
separating the liquid like region from the glassy phase. On the
other hand, if the system is prepared in an energy state above the
free energy threshold, it starts evolving towards the threshold free
energy as the temperature is reduced below $\Tc$. This process slows
down continuously due to the decreasing connectivity of the visited
free energy landscape so that the free energy threshold is never
reached. Therefore, even a vanishingly small driving force is
sufficient in order to stop aging and keep the system sliding over
this threshold free energy \cite{Berthier-Barrat-Kurchan::PRE}.}

\add{This picture motivated a later work, where the existence of a yield
stress could be shown for a $p$-spin system with a finite number of
spins~\cite{Berthier2003}. The work follows the idea that free
energy barriers are finite at finite system size, thus allowing the
thermal activations to play a role which is not possible in the case
of an infinite system. In a system with finite free energy barriers,
thermal activation becomes important since it allows the system to
leave local free energy minima where it is trapped so that deeper
valleys can be reached. A finite force must then be applied to the
system in order to stop this aging process and to impose a steady
sliding motion. For $p\myeq 3$, the finite-sized version of the
model has been investigated by Monte Carlo simulations supporting
the existence of a critical driving force below which the system is
trapped ('solid') and above which it ``flows''
('liquid')~\cite{Berthier2003}.}

A manifest of shear melting of an amorphous solid can be seen in the behavior
of particle displacements. While in the quiescent state a particle is
practically eternally trapped in its local environment (the 'cage'' formed by
its nearest neighbors) it can explore far larger distances as the system
is exposed to an external drive such as a shear motion. Thus, the particles in
a driven glass are not localized but behave as in a fluid where the whole
available space can be explored via diffusive motion.

This ``freedom'' in particle motion is also reflected in the
behavior of the structural correlations. While in the quiescent
system the structural relaxation times continuously grow as the
system is quenched into a glassy phase (aging), eventually exceeding
available experimental time window, all correlations decay to zero
within a finite time in a driven glass.

Based on the above picture of shear-driven melting of an amorphous
solid, one could possibly expect that, regardless of the shear rate,
a finite stress must always be applied in order to enable the system
to overcome free energy barriers so that structural relaxation can
take place. If this conclusion is true, a stress plateau is expected
as soon as the time scale of the inherent structural relaxation is
far beyond the accessible simulation time, since, in this case, the
finite relaxation time in the driven system can be considered as
uniquely imposed by the external drive, a situation very similar to
shear-induced melting.

In order to test this idea, we performed large scale molecular dynamics
simulations of a simple molecular glass (see below for a description of the
model) focusing on the interplay between the structural relaxation and
the rheological response of the system both in the supercooled state
and in the glassy phase. \add{As published in a recent paper \cite{Varnik2006}},
our simulation results clearly support the existence of a
\add{stress plateau at low shear rates} in an amorphous solid.
\add{The focus of that paper was on the stress
response and on a comparison of simulated results with MCT-based theoretical
calculations. Here, we go a step further and study the dynamics of structural
relaxation in the same range of temperatures and shear rates for
which the flow-curves have been studied in reference \cite{Varnik2006}.
Furthermore, we will investigate an interesting consequence of the presence
of a stress plateau (which in this case plays the role of
a yield stress) on a Poiseuille-type flow in a planar channel. It will be
shown that in a region around the channel center where the stress is below the
plateau stress the system behaves as an amorphous solid whereas it
exhibits liquid-like behavior beyond this central region.}

It is noted that the rheological response of exactly the same
model has already been studied in reference
\cite{Berthier-Barrat::JChemPhys2002}. However, due to a rather
restricted range of studied shear rates, no conclusion could be made
on the existence or not of a dynamic yield stress in the glassy
phase. \add{The novelty of our simulations is the extended
range of shear rates allowing a conclusive statement on the
existence of a stress plateau for the present model in the glassy
state. When restricted to the range of shear rates accessible to our
simulations, this stress plateau may be regarded as the yield stress
of the system.}

The paper is organized as follows. After an introduction of the
model and the simulation method in the next section, the effect of
shear on the structural relaxation of an amorphous system will be
discussed in section \ref{section:relaxation}. In particular, it
will be shown that the structural relaxation of an amorphous system
exposed to a shear exhibits signature of the quiescent state so that
a study of the structural relaxation alone is sufficient in order to
find out whether the quiescent state of the (driven) system belongs
to a liquid-like (supercooled) state or to a glassy phase. In
section \ref{section:rheology} the rheological response of the
system is studied. The focus of this section is to show that a
similar information on the quiescent state of the system may also be
gained via a study of the stress response. This is possible since
the stress-shear rate curves of a glass are qualitatively different
compared to those of a supercooled liquid. The information on the
glass transition is, therefore, also encoded in the rheological
response. Section \ref{section:conclusion} compiles our results.
%%%%%%%%%%%%%%%%%%%%%%%%%%%%%%%%%%%%%%
\section{model and simulation method}
\label{section:model} %%%%%%%%%%%%%%%%
A generic glass forming system, consisting of an 80:20 binary mixture
of Lennard-Jones particles (whose types we call A and B) at a total
density of $\rho\myeq \rhoA+\rhoB \myeq 1.2$ and in a cubic box of length $L \myeq 10$
($N \myeq 1200$ particles) is studied \cite{Kob-Andersen::PRL73::1994}.

A and B particles interact via
$U_{\text{LJ}}(r)\myeq
4\epsilon_{\alpha\beta}[(d_{\alpha\beta}/r)^{12}-(d_{\alpha\beta}/r)^6]$,
with $\alpha,\beta\myeq {\text{A,B}}$, $\epsilon_{\text{AB}}\myeq
1.5\epsilon_{\text{AA}}$, $\epsilon_{\text{BB}}\myeq
0.5\epsilon_{\text{AA}}$, $d_{\text{AB}}\myeq 0.8d_{\text{AA}}$,
$d_{\text{BB}}\myeq 0.88d_{\text{AA}}$ and $m_{\text{B}}\myeq
m_{\text{A}}$. The potential was truncated at twice the minimum
position of the LJ potential, $\rcab\myeq 2.245 d_{\alpha\beta}$. The parameters
$\epsilon_{\text{AA}}$, $d_{\text{AA}}$ and $m_{\text{A}}$ define
the units of energy, length and mass.  All other quantities reported
in this paper are expressed as a combination of these units.
The unit of time, for example, is given by
$\tauLJ \myeq d_{\text{AA}}\sqrt{m_{\text{A}} / \epsilon_{\text{AA}}}$ and that of stress
by $\epsilon_{\text{AA}} / d^3_{\text{AA}}$. Equations of
motion are integrated using a discrete time step of $dt \myeq 0.005$.

The system density is kept constant at the value of $1.2$ for all
simulations whose results are reported here. This density is high
enough so that no voids occur at low temperatures and low enough so
that system dynamics remains sensitive to a variation of temperature
(see references \cite{Starr-PRE60-1999,Sastry-PRL85-2000} for
effects of high density/pressure on the liquid-glass transition).

The present model was found suitable for an analysis of many aspects of the mode
coupling theory of the glass transition
\cite{Kob-Andersen::PRE51,Kob-Andersen::PRE52}. In particular, at a
total density of $\rho \myeq 1.2$, equilibrium studies of the model showed
that the growth of the structural relaxation times at low
temperatures could be approximately described by a power law as predicted by the
ideal MCT, $\tau_{\text{relax}}\propto (T-\Tc)^{-\gammamct}$. Here,
$\Tc \myeq 0.435$ is the mode coupling critical temperature of the model
and $\gammamct$ is the critical exponent. For the present binary
Lennard-Jones system, numerical solution of ideal MCT equations
yields a value of $\gammamct \approx 2.5$ \cite{Nauroth-Kob::PRE55}. A
similar value is also obtained for a binary mixture of soft spheres
\cite{Barrat-Latz:JPCM2-1990}.

Simulation results are averaged over 10 independent runs.
For this purpose, ten independent samples are equilibrated at a temperature
of $T \myeq 0.45$ (above $\Tc$) and serve as starting configurations for all simulated
temperatures and shear rates. The temperature is controlled via
Nos\'e-Hoover thermostat \cite{Nose::JCP81,Hoover::PhysRevA31}.
It is set from $T \myeq 0.45$ to the desired value at the beginning of shear, whereby only the
$y$-component of particle velocities is coupled to the heat bath
($x$ being the streaming and $z$ the shear gradient directions,
see also below).

The temperature quench is done only in one step, i.e.\ without a
continuous variation from $T_{\rm{start}}$ to $T_{\rm{end}}$.
However, as the numerical value of $T$ is changed, it takes a time
of the order of the velocity autocorrelation time for the new
temperature to be established. During this period of time the
Maxwell distribution of velocities undergoes changes in order to
adapt itself to a distribution determined by the new temperature.
This time is of order unity (in reduced units) and quite short
compared to all other relevant timescales in the problem.

Previous studies of the stress-strain relation of the same model
showed that the initial transient behavior is limited to strains
below 50\% \cite{Varnik-Bocquet-Barrat::JChemPhys2004}. Indeed, by shifting the time origin in
measurements of various correlation functions, we verified that the
time translation invariance was well satisfied in sheared systems
for strains larger than 50\%. We neglected strains $\gamma \!<\! 100\%$ before
starting the measurements. Unless otherwise stated, all simulations reported
below had a length corresponding to a strain of 7.8 (780\%). In the steady state, correlation
functions were averaged both over independent runs and over
time origins distributed equidistantly along each simulation run.
The shear stress is calculated using the virial expression
\cite{Evans-Morriss::StatMechNonEqLiq1990}
\begin{equation}
\sigma \equiv \sigma_{xz}  \myeq  \frac{-1}{V} \left< \sum_i^{N} m_i
v_{ix}v_{iz} + \frac{1}{2} \sum_{i \neq j}^{N} x_{ij}
F_{zij}\right>, \label{eq:stress}
\end{equation}
where $\langle ... \rangle$ stands for statistical averaging,
$v_i$ and $m_i$ are the velocity and the mass of $i$-th particle, $x_{ij} \myeq x_j-x_i$
and $F_{zij}$ the $z$-component of the force of particle $j$ on $i$.

\begin{figure}
\epsfig{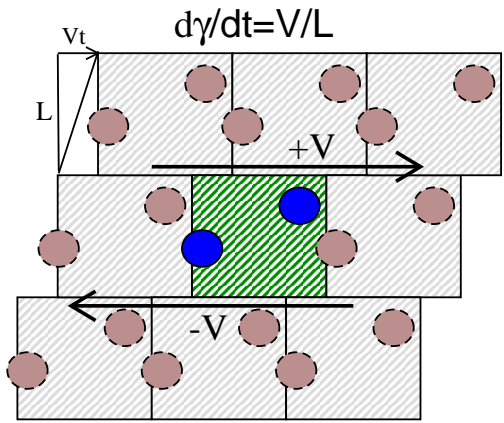}
\caption{An illustration of the Lees-Edwards
boundary condition. The central box depicts a simulation box
containing two particles only. All surrounding cells are images of
the central box. As indicated by bold arrows, the image box on the
top is moved with a constant velocity of V towards right, whereas
the image box on the bottom is moved with the same velocity in the
opposite direction. This defines an overall shear rate of $\gammadot
 \myeq  V/L$, where $L$ is the box length in the vertical direction. Note
that all particles of a given image box have exactly the same
additional velocity $V$ ($-V$). The method is therefore sometimes
referred to as the sliding brick boundary condition. \add{This image is motivated
by a similar figure in reference \cite{Evans-Morriss::StatMechNonEqLiq1990}.}}
\label{fig:slidingbrick}
\end{figure}

\begin{figure}
\epsfig{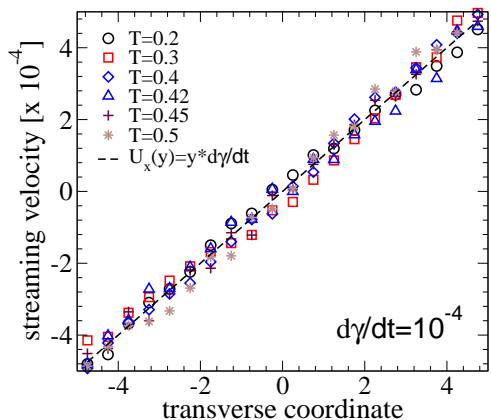}
\caption{Velocity profiles across the system obtained for various temperatures
as indicated (ranging from the glassy phase to the supercooled regime). As
seen here, shear localization is absent at all studied temperatures.
However, at the global shear rate studied here, shear bands may occur
if the shear is imposed via the motion of atomistic walls
\cite{Varnik-Bocquet-Barrat-Berthier::PRL2003}. The use of the SLLOD algorithm in
conjunction with the Lees-Edwards boundary condition is, therefore,
essential in order to ensure a homogeneous shear across the system.}
\label{fig:vprofile}
\end{figure}

Recent studies of the present model in the glassy state showed that
the system may exhibit shear-localization if the shear rate is
imposed by using a conventional Couette cell with moving atomistic
walls~\cite{Varnik-Bocquet-Barrat-Berthier::PRL2003}. A shear
banding is, however, undesired in the context of present analysis
since we are interested in the effects of a homogeneous shear. On
the other hand, simulations of the present model using the so-called
Lees-Edwards boundary conditions along with the SLLOD equations of
motion first proposed by Evans and coworkers
\cite{Evans-Morriss::StatMechNonEqLiq1990} show that a linear
velocity profile forms across the system thus leading to a spatially
constant velocity gradient \cite{Berthier-Barrat::JChemPhys2002}.
We, therefore, also used this approach for our simulations (see
Fig.\ \ref{fig:slidingbrick} for an illustration of the Lees-Edwards
boundary condition). Within this simulation method, we do indeed
observe a spatially constant velocity gradient in all studied cases
(Fig.\ \ref{fig:vprofile}).

The external shear does work on the system. Therefore, in order to
keep the system temperature at a prescribed value, this extra heat
must be removed. This is done by coupling the $y$-component of
particle velocities to the so-called Nos\'e-Hoover thermostat
\cite{Nose::JCP81,Hoover::PhysRevA31}. We impose a flow in
the $x$-direction with a shear gradient in the $z$-direction. The
$y$-component of particle velocities are therefore not affected by
the local flow velocity, thus simplifying the computation of
instantaneous kinetic energy which enters the thermostating part of
equations of motion. This choice also ensures that possible artefact
due to a profile biased thermostat \cite{Evans-Morriss::PRL56-1986}
are absent. Further details about the simulation and the model can
be found in reference \cite{Varnik-Bocquet-Barrat::JChemPhys2004}.

\add{The above choice of the simulation method leads to the
following set of equations of motion,
\begin{eqnarray}
\nonumber
  \dot{x}_i  = p_{xi}/m_i+z_i \gammadot, \; \; \dot{y}_i &=& p_{yi}/m_i,\;\;\;\;\; \dot{z}_i  = p_{zi}/m_i\\
\nonumber
\!\!\dot{p}_{xi} = F_{xi}-p_{zi}\gammadot,\;\;\;\dot{p}_{yi} &=& F_{xi}-\xi p_{yi},\;\; \dot{p}_{zi} = F_{zi}\\
\nonumber
\dot{\xi}= (\sum_i p_{yi}^2/m_i &\!\!\!-\!&\!\! N \kB  T)/Q. \label{eq:motion}
\end{eqnarray}
Here $(x_i,\;y_i,\;z_i)$, $(p_{xi}, \; p_{yi},\;p_{zi})$ and $m_i$
denote coordinates, momenta and the mass of $i$-th particle
respectively. $N\myeq \NA+ \NB$ is the total number of particles,
$\kB$ the Boltzmann constant and $F$ the total force on a particle.
It is noted that the momenta occurring in the above equations are
the peculiar ones, i.e., they correspond to the particles' momenta
in a (local) frame of reference moving with the flow ($\langle p
\rangle=0$) \cite{Evans-Morriss::StatMechNonEqLiq1990}. The
variable, $\xi$, plays the role of a friction (acceleration)
coefficient, since a positive (negative) $\xi$ tends to decrease
(increase) $p_y$. The variation of $\xi$, on the other hand, is
controlled by the deviation of the actual kinetic energy of the
system from that prescribed by the temperature, $T$, of the heat
bath. The parameter $Q$ controls the strength of the coupling  of
the particles' momenta to the heat bath: the smaller Q the stronger
the coupling (see reference \cite{Varnik::Dissertation::Mainz2000}
for more details on thermostating methods).}

%%%%%%%%%%%%%%%%%%%%%%%%%%%%%%%%%%%%%
\section{Structural relaxation in a driven system}
\label{section:relaxation} %%%%%%%%%%%%%%
It is one of the aims of the present section to show that the
structural relaxation of an amorphous system exposed to a shear
already exhibits signature of the quiescent state of the system.
In other words, a study of the structural relaxation in the
driven case allows one to find out whether the quiescent state
of the system belongs to a liquid-like (supercooled) state or
to a glassy phase.

As will be shown below, this possibility of distinguishing
between a sheared glass and a driven supercooled liquid is
closely related to the fact that, in the supercooled state, the
structural relaxation is affected by the external drive
at high shear rates only. As the shear rate tends to
zero, the (shear-independent) equilibrium relaxation is
recovered. In contrast to this, the structural relaxation of
a sheared \emph{glass} is always driven by the imposed shear, since,
in an amorphous solid, there is no equilibrium relaxation in the available
experimental (or simulation) time window.

\begin{figure}
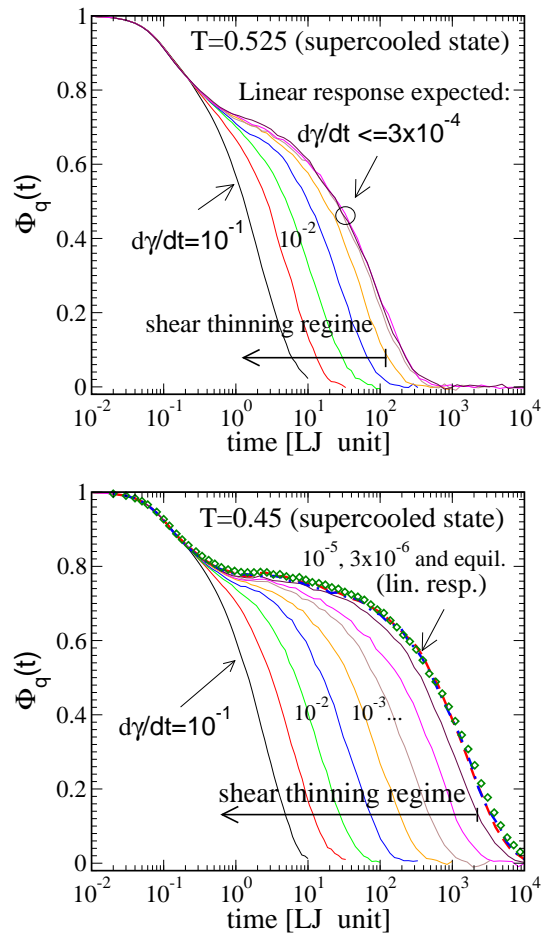

\epsfig{file=gammadot_effect_T0.525.eps,height=60mm,clip=}\vspace*{3mm}
\epsfig{file=gammadot_effect_T0.45.eps,height=60mm,clip=}\vspace*{-2mm}
 \caption[]{Effect of a homogeneous shear
rate on the relaxation behavior of the incoherent scattering
function. $\Phi_q(t)$ is shown for two temperatures, both belonging
to the supercooled regime. The overall behavior is similar at both
temperatures: At high shear rates an acceleration of the system
dynamics is observed, whereas equilibrium curves are recovered at
sufficiently low $\gammadot$. However, a comparison of these two
panels clearly shows that the linear response regime is shifted
towards considerably lower shear rates as temperature is reduced.
The mode coupling critical temperature of the system is $\Tc \myeq
0.435$ \cite{Kob-Andersen::PRE51,Kob-Andersen::PRE52}). From left to
right $\gammadot \myeq 10^{-1},\;3\times
10^{-2},\;10^{-2},...,3\times 10^{-5}$ for $T \myeq 0.525$ and
$\gammadot \myeq 10^{-1},\;3\times 10^{-2},\;10^{-2},...,3\times
10^{-6}$ for $T \myeq 0.45$.} \label{fig:phiq-supercooled}
\end{figure}

\begin{figure}
\epsfig{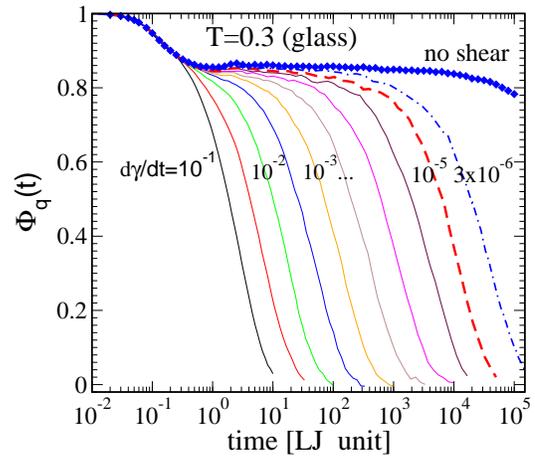}
\caption[]{Same as in Fig.\
\ref{fig:phiq-supercooled} but for a temperature \emph{below} $\Tc$.
The incoherent scattering function, $\Phi_q(t)$, shows a pronounced
two step relaxation. Whereas the short time dynamics is practically
unaffected by shear, the long time behavior of $\Phi_q(t)$ is
clearly dominated by the imposed shear rate. The lower $\gammadot$
the larger $\tau$, the time necessary for a substantial decay of
$\Phi_q(t)$. For comparison, the quiescent case, $\Phi_q(t,\tw; \gammadot \myeq 0)$,
measured after a waiting time of $\tw \myeq 10^5$, is also shown
(see diamonds; in the case of  $\Phi_q(t,\tw; \gammadot \myeq 0)$;
the averaging is done over a large number of initial configurations).
Obviously, after a waiting time of $\tw \myeq 10^5$, the inherent system
dynamics is far slower than the shear driven decay  of $\Phi_q$
for \emph{all} studied shear rates. This must be contrasted to the supercooled
state, where equilibrium behavior is recovered at low $\gammadot$
(Fig.\ \ref{fig:phiq-supercooled}). From left to right,
$\gammadot \myeq 10^{-1},\;3\times 10^{-2},\;10^{-2},...,3\times
10^{-6}$.} \label{fig:phiq-glass}
\end{figure}

In order to illustrate these features, we show in Figs.\
\ref{fig:phiq-supercooled} and \ref{fig:phiq-glass} the incoherent
scattering function ($q_y$ is the $y$-component of the wave vector),
\begin{equation}
\Phi_q(t)  \myeq  \frac{1}{N}\Big< \sum^N_{i=1} \exp[q_y(y_i(t)-y_i(0))]
\Big>, \label{eq:phiq}
\end{equation}
obtained from molecular dynamics simulations at various shear rates
and temperatures. The use of $y$-component naturally eliminates
undesired contributions to particle rearrangements arising, for
example, from the motion of particles with the flow and the
so-called Taylor dispersion (affecting the displacement of particles
along the flow direction as they move in the direction of shear
gradient) \cite{Taylor-ProcRSocLondonA-219::1953}.

We measure $\Phi_q(t)$ for a wave vector of $q \myeq 7.1$, corresponding
to the principal peak of the static structure factor. $\Phi_q(t)$
thus reflects collective particle rearrangements on the scale of
average nearest neighbor distance, the relevant length scale for a
study of the cage effect.

In Fig.\ \ref{fig:phiq-supercooled}, $\Phi_q(t)$ is depicted for two
temperatures from the supercooled regime. In each panel, the
incoherent scattering function is shown for various shear rates
ranging from a regime, where the system dynamics is accelerated by
the imposed shear, to lower $\gammadot$ where the structural relaxation is independent of
shear rate. First note that,
at short times, all curves overlap regardless of the imposed shear rate indicating that
the short time dynamics is unaffected by the shear. The situation
is, however, different when long time behavior of $\Phi_q(t)$ is
considered. At high $\gammadot$, the final relaxation of $\Phi_q(t)$
strongly depends on the magnitude of the imposed shear rate: The
higher $\gammadot$ the faster the decay of $\Phi_q(t)$. However, as
$\gammadot$ is progressively decreased, the dependence of
$\Phi_q(t)$ on $\gammadot$ becomes weaker, eventually disappearing
for the lowest shear rates shown in the figure.

A comparison of $\Phi_q(t)$ for a shear rate of $\gammadot  \myeq 10^{-4}$
in both panels of Fig.\ \ref{fig:phiq-supercooled} shows that a
shear rate which does not affect the equilibrium relaxation at a high
temperature (here, $T \myeq 0.525$) may induce an acceleration of the system
dynamics at a lower temperature ($T \myeq 0.45$).
Interestingly, as will be shown in the next section  (Fig.\ \ref{fig:stress}),
the stress response exhibits a  linear (non linear) behavior for roughly the
same shear rates where $\Phi_q(t)$ is unaffected (affected) by shear.

\add{The above comparison follows a similar plot for $T=0.5$ in reference
\cite{Berthier-Barrat::JChemPhys2002} (see the upper left panel in Fig.\ 8).
The new aspect in our plots is the study of considerably wider
range of shear rates allowing to decrease $\gammadot$ sufficiently in order
to recover equilibrium behavior at a temperature as
low as $T=0.45$, a feature suggested but not shown in
reference \cite{Berthier-Barrat::JChemPhys2002}.}

The discussion of Fig. \ref{fig:phiq-supercooled} underlines the
idea that the linear or non linear nature of the system response to
an imposed shear rate is the outcome of the competition between the
time scale imposed by the flow and the inherent (structural)
relaxation time of the system. The faster the inherent system
dynamics, the larger the range of shear rates in the linear regime
\cite{Berthier-Barrat-Kurchan::PRE,Fuchs-Cates::FaradayDiscuss-2003,Miyazaki-Reichman-PRE66}.

Figure \ref{fig:phiq-glass} illustrates the effect of shear on the
relaxation behavior of a glass. For this purpose, $\Phi_q(t)$ is
depicted for a temperature far below the mode coupling critical
temperature of the model. Again, at short times, all curves are
identical irrespective of the imposed shear rate. As for the long
time behavior, $\Phi_q(t)$ exhibits a dependence on $\gammadot$ at
all simulated shear rates. \add{The plot in Fig.\ \ref{fig:phiq-glass} 
is also motivated by the upper right panel in Fig.\ 8 of reference 
\cite{Berthier-Barrat::JChemPhys2002}. In addition to a wider range of 
shear rates shown in our plot, Fig.\ \ref{fig:phiq-glass} also compares 
the sheared $\Phi_q(t)$ to its quiescent counterpart,
$\Phi_q(t,\tw;\gammadot=0)$, measured after a waiting time of $\tw=10^5$ (and
averaged over independent initial samples). This allows to
unambiguously demonstrate the fact that, even at the lowest 
studied shear rate ($\gammadot=3\times 10^{-6}$), the final decay of the relaxation function at 
$T=0.3$ is indeed a result of imposed shear.
}

In the linear response regime, transport coefficients can often be
expressed as time integrals of equilibrium correlation functions.
The shear viscosity, for example, is given by the well-known
Green-Kubo relation $\eta \myeq \int_0^{\infty} g(t) dt $, where
$g(t) \myeq V/(\kB T) \langle \sigma_{xy}(t) \sigma_{xy}(0) \rangle$
($V$ denotes the system volume and $\sigma_{xy}$ a component of the
stress tensor with zero mean, $\langle \sigma_{xy}\rangle=0$). Thus,
in a range where $g(t)$ becomes independent of the applied shear
rate, the shear viscosity must also be a constant (Newtonian behavior).

\begin{figure}
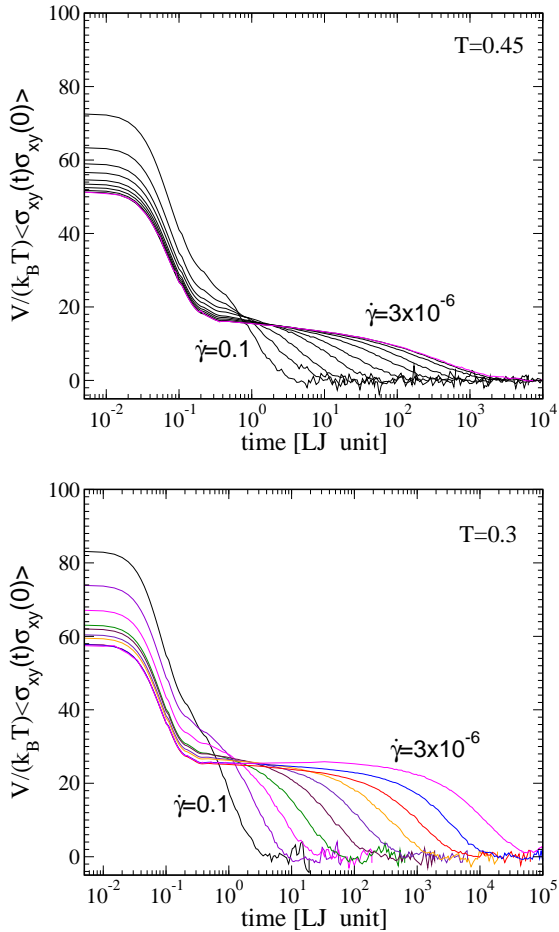

\epsfig{file=stress-autocorr-T0.45.eps,height=60mm,clip=}\vspace*{3mm}
\epsfig{file=stress-autocorr-T0.3.eps,height=60mm,clip=}\vspace*{-2mm}
 \caption[]{Effect of a homogeneous shear rate on the relaxation behavior
of the stress autocorrelation function,
$\langle\sigma_{xy}(t)\sigma_{xy}(0)\rangle$
for a temperature of $T=0.45$ (supercooled state) and $T=0.3$ (glassy phase).
In the supercooled state, an acceleration of the stress relaxation
is observed at high shear rates, whereas equilibrium curves
are recovered at sufficiently low $\gammadot$. In the glass,
 on the other hand, no equilibrium curve exists within the
simulation time window so that even the smallest shear rate studied
significantly alters the system dynamics. From left to right
$\gammadot \myeq 10^{-1},\;3\times
10^{-2},\;10^{-2},...,3\times 10^{-6}$. Note that, contrary to $\sigma_{xz}$
whose average gives the shear stress in the system, $\langle \sigma_{xy} \rangle =0$.
This explains why the studied correlation function decays to zero and not to a finite positive value.
}
\label{fig:stress-autocorr}
\end{figure}

As can be seen from Eq.\ (\ref{eq:stress}), the shear stress can be
expressed as a function of particle positions and (much faster
variables) momenta and forces, whereas $\Phi_q(t)$ depends on the
particle positions only. Thus, at least for wave vectors
corresponding to the average interparticle distance, one expects a
faster decay of stress autocorrelation function compared to a
relaxation of $\Phi_q(t)$. \add{This expectation is born out in
Fig.\ \ref{fig:stress-autocorr}. However, despite the fact that the
time scale for the final decay of $g(t)$ is roughly by one order of
magnitude shorter than the decay time of $\Phi_q(t)$ (for $q$ being
the maximum of the structure factor), the cross over from the shear
thinning behavior to equilibrium relaxation occurs at practically
the same $\gammadot$ for the both types of correlation functions
(compare the data for $T=0.45$ in Figs.\ \ref{fig:phiq-supercooled}
and \ref{fig:stress-autocorr}).}

\add{This observation underlines the relevance of $\Phi_q(t)$ as an
appropriate quantity for at least qualitative studies of non-linear
rheology. In a $\gammadot$-range where $\Phi_q(t)$ becomes
independent of the shear rate, the shear viscosity is also expected
to be independent of $\gammadot$. The study of the flow curves
presented below (Fig. \ref{fig:stress}) confirms this expectation.}

We examine the validity of the so-called "time-shear
superposition principle", predicted both within the spin glass theory
\cite{Berthier-Barrat-Kurchan::PRE} and the non-equilibrium
MCT \cite{Fuchs-Cates::FaradayDiscuss-2003}.
This property indicates that the shape of
$\Phi_q(t, \gammadot)$ at sufficiently large times is independent of
shear rate. More precisely, $\Phi_q(t,\gammadot) \myeq F_q(\hat{t})$,
where $\hat{t} \myeq t/\tau(\gammadot)$ is a dimensionless time. The shear
rate dependent time, $\tau(\gammadot)$, characterizes the time scale
of the final decay of the correlator (the so-called
$\alpha$-relaxation).

\begin{figure}
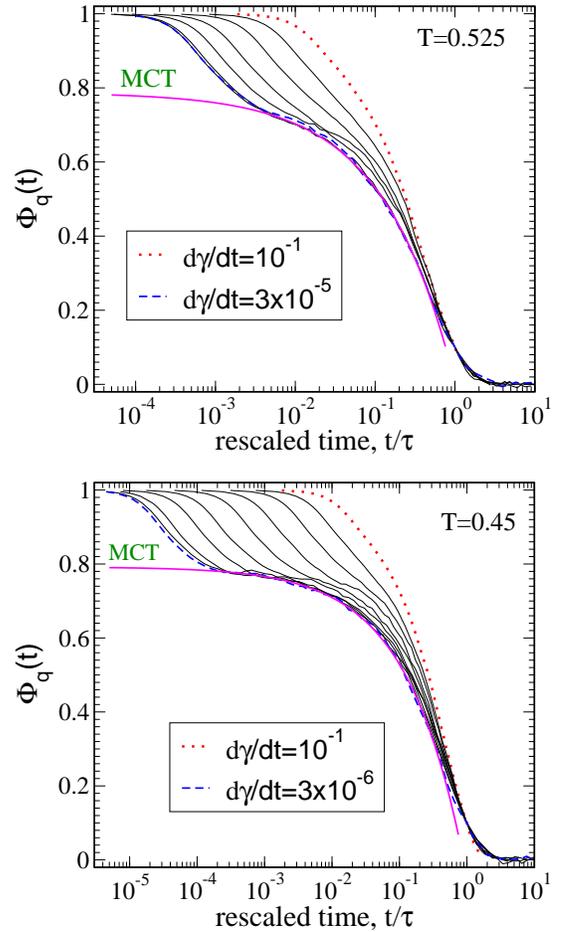

\epsfig{file=time_shear_superposition_T0.525.eps ,height=60mm,clip=}\vspace*{3mm}
\epsfig{file=time_shear_superposition_T0.45.eps,height=60mm,clip=}\vspace*{-2mm}
\caption{A test of time-shear
superposition principle. Same data as in Fig.\
\ref{fig:phiq-supercooled} are plotted versus rescaled time
$t/\tau(\gammadot)$. At low $\gammadot$ and for long times, curves
follow a master curve whereas deviations from the master curve are
observed for higher shear rates. The thick solid line is a fit using
the von Schweidler law, Eq.\ (\ref{eq:vonSchweidler}), with
$f_q \myeq 0.79$, $b \myeq 0.49$ and $A \myeq 0.783$ for $T \myeq 0.525$ and $f_q \myeq 0.79$,
$b \myeq 0.53$ and $A \myeq 0.922$ for $T \myeq 0.45$. Note that the von Schweidler
law is predicted within equilibrium MCT as an approximation to the
late $\beta$-relaxation (see Refs.\ \cite{Kob-Andersen::PRE51,Kob-Andersen::PRE52} for a more
detailed test of ideal MCT predictions of the von Schweidler law).}
\label{fig:Fq:sup}
\end{figure}

This property is checked for in Fig.\ \ref{fig:Fq:sup}. In this
figure, the same data as shown in Fig.\ \ref{fig:phiq-supercooled} are
depicted versus rescaled time $t/\tau(\gammadot)$ ($\tau$ is
estimated from $\Phi_q(t \myeq \tau) \myeq 0.1$). As seen from the both panels of
Fig.\ \ref{fig:Fq:sup}, there is a group of curves following a
master curve. These curves belong to low $\gammadot$, whereas at
higher $\gammadot$ the time-shear superposition principle is
violated. This is reminiscent of similar deviations observed in the 
non-driven supercooled state where the role of $\gammadot$ is played by
 temperature \cite{Kob-Andersen::PRE52,Bennemann-Paul-Binder-Duenweg::PRE57}.

For the both temperatures shown in Fig.\ \ref{fig:Fq:sup}, there is an
intermediate time window (``intermediate'' in the sense that it is long compared to
the initial decay time but short compared to the final decay time of
$\Phi_q(t)$), known as MCT-$\beta-$relaxation regime,  for which the
master curve can be well fitted to the so-called von Schweidler law,
\begin{equation}
\Phi_q(t) = f_q - A(t/\tau)^b \label{eq:vonSchweidler}.
\end{equation}
Here, the non ergodicity parameter, $f_q$, determines the height of
the plateau (to which the short time decay leads) and $b$ is the
so-called von Schweidler exponent. We find practically the
same numerical value for $f_q$ and very close values for $b$ at both
temperatures. This is consistent with ideal MCT, which yields a
theoretical justification of this empirical law
\cite{Goetze::LesHouches::1989}. Indeed, here we use the
results of \emph{equilibrium} MCT, since at low shear rates (where
the theory is supposed to apply) $\Phi_q(t)$ recovers its
equilibrium behavior.

\begin{figure}
\epsfig{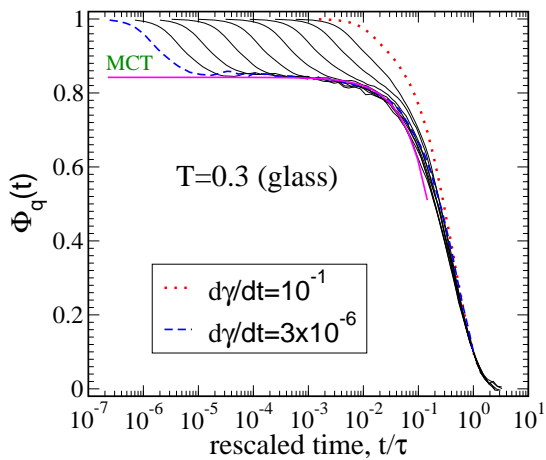}
\caption{ A test of time-shear
superposition principle in the glass. Data in Fig.\
\ref{fig:phiq-glass} are plotted versus rescaled time
$t/\tau(\gammadot)$. At low $\gammadot$ and for long times, curves
follow a master curve whereas deviations from the master curve are
observed for higher shear rates. The thick solid line is a fit to
Eq.\ (\ref{eq:vonSchweidler}) with an exponent of $b \myeq 1$
\cite{Fuchs-Cates::FaradayDiscuss-2003}. This gives a non ergodicity
parameter, $f_q \myeq 0.84$, which is larger than the one obtained
in the supercooled state.} \label{fig:Fq:glass}
\end{figure}

For the glass, a similar plot as in Fig.\ \ref{fig:Fq:sup} is shown
in Fig.\ \ref{fig:Fq:glass}. Obviously, also in the glass, the low
shear rate data confirm the validity of the time-shear superposition
principle. It is worth mentioning that, Eq.\
(\ref{eq:vonSchweidler}) remains valid also in the case of a driven amorphous
solid. However, in contrast to the case of the quiescent system, where
the von Schweidler exponent, $b$, may depend on the system properties at equilibrium,
the non-equilibrium MCT predicts $b \myeq 1$ for the case of a driven
glass \cite{Fuchs-Cates::PRL-2002,Fuchs-Cates::FaradayDiscuss-2003}. 
Using $b\myeq 1$, a fit to Eq.\ (\ref{eq:vonSchweidler}) is also shown
in Fig.\ \ref{fig:Fq:glass}. As also expected from the enhanced solid-like character
of the system at $T\myeq 0.3$ compared to $T\myeq 0.45$ and $T\myeq 0.525$,
the value obtained for the non ergodicity parameter, $f_q \myeq 0.84$, is
significantly larger than  $f_q \myeq 0.79$, obtained in the supercooled state
(Fig.\ \ref{fig:Fq:sup}).

%%%%%%%%%%%%%%%%%%%%%%%%%%%%%%%%%%%%%
\section{Rheological response}
\label{section:rheology} %%%%%%%%%%%%%%
In this section, the rheological response of the system to a
spatially constant (homogeneous) shear rate is studied. It will be shown that the
stress response in the supercooled state is \emph{qualitatively}
different from that in the glass. In the supercooled regime, the non
linear response at high shear rates is always followed by a linear
relation between the shear stress and the shear rate (resulting in a
constant viscosity) as the shear rate is sufficiently lowered.
However, as temperature is decreased, the linear regime is shifted
towards progressively lower shear rates and eventually disappears in
the glassy phase. In other words, in the glass, any shear rate, as
low as it might be, leads to a non linear response.

In order to elucidate this aspect, we investigate the behavior of
the flow curves (shear stress versus shear rate) across an
ideal glass transition.

\begin{figure}
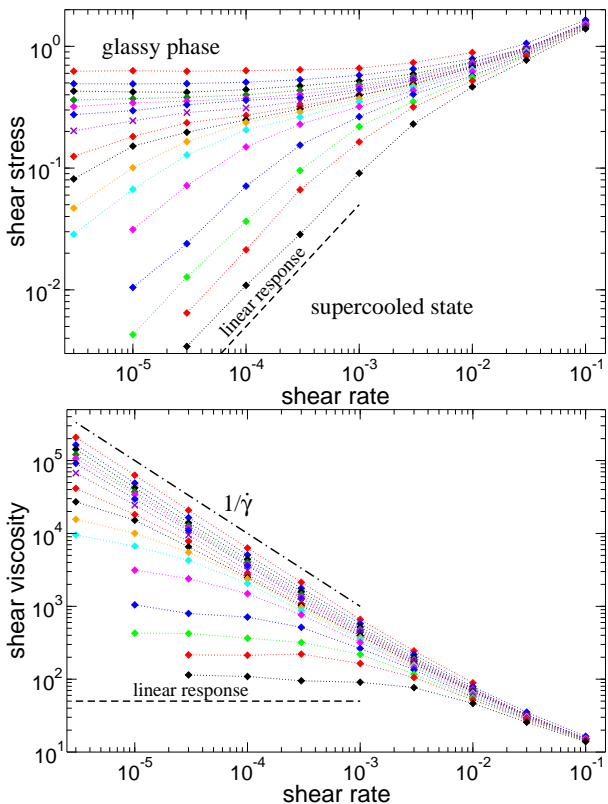

\epsfig{file=stress-simulation-only-DT0.005.eps,width=80mm, clip=}
\epsfig{file=viscosity-simulation-only-DT0.005.eps,width=80mm, clip=}
\caption[]{Top: Simulated shear stress versus shear rate for
various temperatures ranging from the glassy state to the supercooled
regime (from top to bottom:
$T \myeq 0.01,\; 0.2,\; 0.3,\; 0.34,\; 0.36,\; 0.38,\; 0.4,\; 0.42,\; 0.43,\; 0.44,$
$0.45,\; 0.47,\; 0.5,\; 0.525,\;
0.55,\; 0.6$). In the supercooled state, the shear thinning behavior at
high shear rates crosses over to the linear response as $\gammadot
\to 0$. With decreasing temperature, the linear response is,
however, reached at progressively lower $\gammadot$. In the glass,
on the other hand, the shear stress becomes independent of
$\gammadot$ at low shear rates (stress plateau). \add{Figure adapted from reference 
\cite{Varnik2006}.}
Bottom: Apparent shear
viscosity, $\eta_{\text app}\equiv \sigma/\gammadot$ for exactly the same
temperatures as shown in the case of the shear stress.  Not unexpectedly, for
$\gammadot\to 0$, $\eta_{\text app}$ reaches a constant value in the
supercooled state, whereas it grows as $1/\gammadot$ in the glass.}
\label{fig:stress}
\end{figure}

\begin{figure}
\epsfig{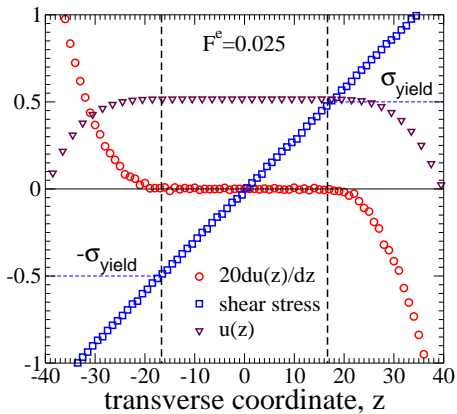}
\caption[]{Flow through a three dimensional planar channel
of the model fluid studied in this paper.
The flow is generated by imposing
a constant body force of $\Fe=0.025$ on each particle. At a
temperature of $T=0.2$ and a total density of $\rho=1.2$, the model
exhibits a stress plateau at low shear rates which, for the present purpose, 
plays the role of a yield stress, $\sigmay \approx 0.5$
(see Fig.\ \ref{fig:stress}). One can easily show that the stress in
the channel behaves as $\sigma=\rho \Fe z$, where $z$ is the
transverse coordinate ($z=0$ being the channel center)
\cite{Varnik-Binder::JChemPhys2002}.
Obviously, for $ |z| \le 16.67 ( \approx \sigmay/(\rho\Fe))$,
the stress in the channel is below the yield stress. One thus
expects the system to behave as a solid body for $|z| \le 16.67$: Either it
should be at rest or move with a constant velocity (zero velocity
gradient). Indeed, an inspection of the velocity profile, $u(z)$, and its
derivative (rescaled to fit into the figure) confirms this
expectation (the region delimited by the two vertical dashed lines).
For $|z|>16.67$, on the other hand, the local shear stress exceeds the
yield stress of the model system thus leading to its liquid like
behavior. The system flows with a shear rate which non-linearly
increases upon increasing stress. Note that if in the present case
the channel width becomes smaller than $\approx  33.34$ (Lennard-Jones units),
there will be no region with stresses above $\sigmay$. As a
consequence, there will be no liquid-like region. The whole system
will behave like a solid body sticking to the boundaries of the
channel (flow blockage). This expectation is confirmed by our
simulations (not shown here).}
\label{fig:yieldstressLJ}
\end{figure}

Simulation results on the stress response to a time independent shear
(flow-curves) are shown in the top panel of Fig.\ \ref{fig:stress}
for temperatures ranging from the supercooled state to the glassy
phase. In the supercooled state, the shear thinning behavior at
high shear rates crosses over to the linear response as $\gammadot
\to 0$. However, as the system is cooled toward the glassy phase,
the linear response is reached at progressively lower shear rates.
In the glass, on the other hand, the shear stress becomes independent of
$\gammadot$ at low shear rates. \add{For many practical purposes (see
e.g.\ Fig.\ \ref{fig:yieldstressLJ}) this stress plateau plays the role
of a yield stress: If the system is subject to stresses
 below this plateau stress, it behaves like a solid body 
(zero velocity gradient) while for higher stresses it exhibits a
liquid-like character.}  Let us mention that a stress plateau is also
observed in recent experiments on the rheology of
dense colloidal dispersions \cite{Petekidis2004,Fuchs2005}.

The bottom panel of Fig.\ \ref{fig:stress} depicts the same data as
in the top panel in a different way. Here, the apparent shear
viscosity, $\etaapp\equiv \sigma /\gammadot$, is plotted against
$\gammadot$. The presence of a linear response regime in the
supercooled state is now illustrated as a constant viscosity line,
whereas the apparent $1/\gammadot$-divergence of the viscosity in the glass (for
$\gammadot \to 0$) is also nicely born out. From this panel, one can
easily extract, for temperatures in the supercooled state, the shear
rate at which the crossover from linear to non linear response takes
place. This gives $\gammadot \myeq 3\times 10^{-4}$ for $T \myeq 0.525$ and
$\gammadot \myeq 3\times 10^{-5}$ for $T \myeq 0.45$. Interestingly, these values
also mark the beginning of the shear rate dependence of $\Phi_q(t)$
for the corresponding temperatures (Fig.\ \ref{fig:phiq-supercooled}).

\add{Let us work out an interesting consequence of the presence of a yield
stress on the flow behavior of the system. For this purpose, we simulate
a Poiseuille type flow in a planar channel. The study of such a situation
is interesting since the stress in a Poiseuille-type flow is zero
in the channel center and increases linearly with the distance from it.
If the fluid under consideration exhibits (at least within the time
window accessible for the simulation) a finite yield stress,
one may expect that the fluid portion in a certain region around
the channel center (defined by the condition that the stress
in this part of the channel be lower than the yield stress)
should behave like a solid body while it should flow 
like a liquid further away from this region.}

\add{The above discussed Poiseuille-type simulation is performed in
the following way. We first equilibrate a system of size
$L_x\times L_y\times L_z=30 \times 30 \times 86$ (containing 92880
particles) at a temperature of $T=0.45$ and then quench it to a
temperature of $T=0.2$. The system is then aged for a waiting time
of $\tw=10^4$ in order to suppress the fast aging process which
occurs directly after the $T$-quench and allow the system to solidify. 
After this period of time, two
solid walls are introduced by immobilizing all particles whose
$z$-coordinate satisfies $|z|>40$ (note that $z\in [-43\; 43]$
in general; this gives rise to walls of three particle diameter
thickness). A flow is then imposed by applying on each particle a
constant force of $\Fe=0.025$ (in LJ units). In the case of a liquid
at equilibrium, this would give rise to a parabolic velocity
profile. However, as Fig.\ \ref{fig:yieldstressLJ} clearly
demonstrates, the velocity profile rather exhibits a behavior
expected for a yield stress fluid: In the central region defined by
the condition of $\sigma \le \sigmay \approx 0.5$ 
(see the stress plateau at $T=0.2$ in the upper panel of 
Fig.\ \ref{fig:stress}) the velocity profile
is flat with a zero gradient while it gradually departs from this
constant behavior (shear rate becomming nonzero) beyond this central
part of the channel. An interesting consequence of this behavior is
that, if for a given driving force (pressure gradient) the channel
width is too small to allow the formation of stresses above the
system's yield stress, no flow will occur and the whole system will
behave like a solid body.}

%%%%%%%%%%%%%%%%%%%%%%%%%%%%%%%%%%%%%%%
\section{Summary}
\label{section:conclusion}
%%%%%%%%%%%%%
We report results of large scale molecular dynamics simulations on the
structural relaxation and the rheological response of a well
established glass forming model, an 80:20 binary Lennard-Jones mixture
first introduced by Kob and Andersen
in the context of the dynamics  of supercooled liquids \cite{Kob-Andersen::PRL73::1994}.

Previous studies of the present model
\cite{Kob-Andersen::PRL73::1994,Kob-Andersen::PRE51,Kob-Andersen::PRE52}
showed that the model was suitable for an analysis of many aspects of the so-called mode
coupling theory of the glass transition (MCT)
\cite{Goetze::LesHouches::1989,Goetze-Sjoegren::RepProgPhys55}. In
particular, for a total density of $\rho \myeq 1.2$,
an ideal computer glass transition (in the sense that the
relaxation times approximately obey a power law divergence predicted by
ideal MCT) was observed at a mode coupling critical temperature of
$\Tc \myeq 0.435$.

The present work is strongly motivated by recent theoretical progress on
the field of the rheology of dense amorphous systems. While various
theoretical approaches
\cite{Sollich::PRE-1998,Berthier-Barrat-Kurchan::PRE,Fuchs-Cates::FaradayDiscuss-2003}
provide a coherent description of the rheological
response of a supercooled liquid (prediction of shear thinning at high shear
rates (or driving force), $\gammadot$, followed by linear response as
$\gammadot$ approaches zero), they make different predictions
with regard to the stress response of an amorphous solid at low shear rates.

In particular, the issue of dynamic yield stress, $\sigmay$,
(shear stress in the limit of $\gammadot \to 0$) remains
controversial. While approaches based on the numerical studies of
disordered $p$-spin systems \add{in the thermodynamic limit} 
\cite{Berthier-Barrat-Kurchan::PRE} suggest that $\sigmay$ identically vanishes in the glassy phase,
both the semi-phenomenological ``soft glassy rheology'' model
\cite{Sollich::PRE-1998} (SGR, an extension of the trap model
\cite{Bouchaud::JPhysI-1992} taking into account the presence of an
external drive) as well as the idealized version of the
non-equilibrium MCT proposed in reference
\cite{Fuchs-Cates::PRL-2002} predict the existence of a finite
dynamic yield stress in the glassy phase.

Nevertheless, qualitatively similar statements on the relaxation
behavior of the correlation functions in a driven disordered solid
are made both within the $p$-spin model and in the non-equilibrium
MCT. In particular, shear melting of a glass and the recovery of the
time translation invariance (shear stops aging
\cite{Kurchan::condmat1998-2000}) is commonly predicted by both
these approaches.

\add{An explanation for the absence of dynamic yield stress
in the $p$-spin model is given in reference
\cite{Berthier-Barrat-Kurchan::PRE} relating this property
to the presence of infinite free energy barriers
occurring in the glassy phase in the thermodynamic limit
($N\to \infty$). This picture has been checked by studying via Monte
Carlo simulations a $p$-spin model with a finite number of spins,
whereby considering a case with \emph{finite} free energy barriers.
Indeed, a yield stress in the glass is observed
within these studies~\cite{Berthier2003}.}

\add{The effect of shear on the relaxation behavior of the intermediate 
scattering function, $\Phi_q(t)$, 
as well as the stress autocorrelation function, $g(t)$, 
is studied both in the supercooled state and in the glass. 
In the supercooled state, high shear rates lead to an accelerated decay
of the correlation function (shear thinning)  compared to the 
equilibrium relaxation which is recovered at sufficiently low 
shear rates.}

\add{Interestingly, despite the fact that $g(t)$ decays much faster than
$\Phi_q(t)$ (for $q$ corresponding to the inverse of the nearest neighbor distance), 
the cross over shear rate from shear thinning to equilibrium
relaxation is practically the same for the both types of the correlation 
functions $\Phi_q(t)$ and $g(t)$ 
(compare the data for $T=0.45$ in Figs.\ \ref{fig:phiq-supercooled}
and \ref{fig:stress-autocorr}). Furthermore, this cross over shear rate
also marks the change from linear to non-linear behavior in the 
stress response (Fig.\ \ref{fig:stress}). These observations underline 
the role of the intermediate scattering function as an appropriate observable 
for a study of at least qualitative features of non-linear rheology.}

\add{In the glass, on the other hand, our simulations clearly 
indicate that the dynamic yielding (shear melting) of an amorphous 
solid (Fig.\ \ref{fig:Fq:glass}) is accompanied by the presence of a finite stress plateau
(Fig.\ \ref{fig:stress}), which we identify as the yield stress
of the model given the restriction to the accessible
observation time window.} It is worth mentioning that a
stress plateau is also found in recent experiments on the rheology of
dense colloidal dispersions \cite{Petekidis2004,Fuchs2005}.

\add{An interesting consequence of the presence of a yield stress is 
discussed for the case of a flow driven by an external body force
(equivalent of a pressure driven Poiseuille-type channel flow; 
see Fig.\ \ref{fig:yieldstressLJ}). This choice is motivated by the fact
that in such a situation the stress in the channel center is zero 
and linearly increases
with distance from the middle of the channel. One thus expects
a solid-like behavior in a central region (defined by $\sigma(z)<\sigmay$; $z$
denoting the transverse coordinate measured from the channel center)
and a fluid-like behavior in the region between this central part and the
channel walls ($\sigma(z)>\sigmay$). As shown in Fig.\
\ref{fig:yieldstressLJ}, this expectation is nicely born out.
An interesting consequence of
this behavior is that, if for a given driving force (pressure gradient)
the channel width is too small to allow the formation of stresses above
$\sigmay$, the whole system will behave
like a solid body sticking to the walls.}

\section*{Acknowledgments}
We thank M. Fuchs, M.E. Cates and J.-L. Barrat for useful
discussions. Many thanks to J. Horbach for careful reading of this
manuscript and his instructive comments. 
During this work, F.V. was supported by the Deutsche Forschungsgemeinschaft
(DFG), Grant No VA 205/3-2. Generous grants of simulation time by the
ZDV-Mainz, PSMN-Lyon and IDRIS (project No 031668-CP: 9) are also acknowledged.

\end{document}